\documentclass[%
 preprint,
 amsmath,amssymb,
 prx]{revtex4-2}

\DeclareMathOperator{\Tr}{Tr}
\DeclareUnicodeCharacter{2009}{\,} 
\usepackage{graphicx}
\usepackage{dcolumn}
\usepackage{bm}
\usepackage{lipsum}
\usepackage{siunitx}

\begin{document}

\title{High Fidelity 12-Mode Quantum Photonic Processor Operating at InGaAs Quantum Dot Wavelength}

\author{Michiel de Goede}
\email{m.degoede@quixquantum.com}
%\affiliation{%
% QuiX Quantum B.V., 7521 AN Enschede, The Netherlands
%}%
\author{Henk Snijders}
\affiliation{%
 QuiX Quantum B.V., 7521 AN Enschede, The Netherlands
}%
\author{Pim Venderbosch}
\affiliation{%
 QuiX Quantum B.V., 7521 AN Enschede, The Netherlands
}%
\author{Ben Kassenberg}
\affiliation{%
 QuiX Quantum B.V., 7521 AN Enschede, The Netherlands
}%
\author{Narasimhan Kannan}
\affiliation{%
 QuiX Quantum B.V., 7521 AN Enschede, The Netherlands
}%
\author{Devin H. Smith}
\affiliation{%
 QuiX Quantum B.V., 7521 AN Enschede, The Netherlands
}%
\author{Caterina Taballione}
\affiliation{%
 QuiX Quantum B.V., 7521 AN Enschede, The Netherlands
}%
\author{Jörn P. Epping}
\affiliation{%
 QuiX Quantum B.V., 7521 AN Enschede, The Netherlands
}%
\author{Hans van den Vlekkert}
\affiliation{%
 QuiX Quantum B.V., 7521 AN Enschede, The Netherlands
}%
\author{Jelmer J. Renema}
\affiliation{%
 QuiX Quantum B.V., 7521 AN Enschede, The Netherlands
}%

\date{\today}

\begin{abstract}
Reconfigurable quantum photonic processors are an essential technology for photonic quantum computing. Although most large-scale reconfigurable quantum photonic processors were demonstrated at the telecommunications C band around 1550\,nm, high-performance single photon light sources utilizing quantum dots that are well-suited for photonic quantum computing operate at a variety of wavelengths. Thus, a demand exists for the compatibility of quantum photonic processors with a larger wavelength range. Silicon nitride (SiN) has a high confinement and wide transparency window, enabling compact, low-loss quantum photonic processors at wavelengths outside the C band. Here, we report a SiN universal 12-mode quantum photonic processor with optimal operation at a wavelength of \SI{940}{nm}, which is compatible with InGaAs quantum dot light sources that emit light in the \SIrange{900}{970}{nm} wavelength range. The processor can implement arbitrary unitary transformations on its 12 input modes with a fidelity of 98.6\%, with a mean optical loss of \SI{3.4}{dB/mode}.
\end{abstract}
\maketitle

\section{\label{sec:introduction}Introduction}

Photonics is one of the most promising approaches to quantum computing, as it is one of the only two platforms to have demonstrated a quantum advantage so far \cite{zhong_quantum_2020}, the other being based on superconducting qubits \cite{Arute_Google_2019}. Photonic quantum computing benefits from the use of quantum states that both have a low decoherence, due to their weak interaction with the environment, and preserve their coherence at room temperature. Linear optics is the backbone of photonic quantum computing due to its simple generation of photon entanglement via quantum interference on a normal beamsplitter. A quantum photonic processor, i.e., a linear optical interferometer, is therefore the central component of a photonic quantum computer. A photonic processor requires universal reconfigurability to implement arbitrary transformations on the input optical states, low losses to preserve the quantum information, and must be large in scale to obtain the necessary complexity of the quantum computation. Integrated photonics meets these conditions and offers a scalable, mature, and commercially available tool for large-scale, phase-stable, and universally reconfigurable linear optical interferometers \cite{Wang_processor_2020}.

Photonic quantum computing requires the integration of a photonic processor with a single photon light source that has both a high brightness and high indistuigshability. Quantum dot (QD) single photon sources are a mature technology that produces highly indistinguishable photons at a large rate \cite{Somaschi_QD_2016, Xing_QD_2016}. Currently, single-photon sources based on QDs can be reproducibly realized with an average indistinguishability of 90\% and brightness---or photon emission probability per pulse---of 13\% for single photon emission \cite{Ollivier_QD_2020}. Furthermore, QDs can be integrated on chip in micropillars \cite{Somaschi_QD_2016} or photonic crystals \cite{Uppu_QD_2020}. State of the art QDs operate in the \SIrange{900}{970}{nm} wavelength range. High quality QDs operating at longer wavelengths are much harder to produce because the physical dimensions of the QDs increases and they are then much more subject to material stress and impurities \cite{Senellart_QD_2017}. QD sources have been used for a wide variety of quantum information processing experiments such as Boson Sampling \cite{Loredo_BS_2017, Wang_BS_2017}, cluster state generation \cite{Istrati_QD_19, Schwartz_QD_2016} and quantum networks \cite{Lu_QD_2021}. Despite these excellent properties, no photonic processors have been demonstrated that are compatible with the wavelength range of InGaAs QDs.

Various quantum photonic processors were previously demonstrated that operate at wavelengths in the telecommunications C band (\SIrange{1530}{1565}{nm}) based on silicon-on-insulator (SOI) \cite{harris_quantum_2017} or silicon nitride (SiN) \cite{taballione_universal_2021}. In fact, the largest universal processor, supporting 20 modes was recently demonstrated on the SiN material platform with low-losses and a high degree of reconfigurability \cite{Taballione_20mode_2022}. However, there is a lack of large-scale, fully reconfigurable quantum photonic processors operating at wavelengths outside the C band. There are some examples of processors capable of photonic quantum information processing in the \SIrange{750}{850}{nm} wavelength range, but they were either not universally reconfigurable \cite{crespi_integrated_2013, Dong_processor_2022, Hoch_processor_2021, tillmann_experimental_2013, spring_boson_2013}, or small-scale with up to 6 modes \cite{carolan_universal_2015, mennea_modular_2018}. For realizing large-scale, low-loss photonic processors at a wide range of operating wavelengths, SiN is the most promising material due to its high optical confinement and integration density \cite{taballione_88_2019} and wide transparency window ranging from \SIrange{400}{2350}{nm} \cite{c_g_h_roeloffzen_low-loss_2018}, which is compatible with the emission wavelength of InGaAs QD single photon sources as well as most other candidate single-photon emitters \cite{eisaman2011invited}. 

In this paper, we report a SiN high-fidelity, low-loss and universal quantum photonic processor designed for operation at \SI{940}{nm}. The demonstrated operating wavelength allows integration with InGaAs quantum dots for a scalable approach towards photonic quantum computing.

\section{\label{sec:chip} The Quantum Photonic Processor}

\begin{figure}[tbh]
  \includegraphics{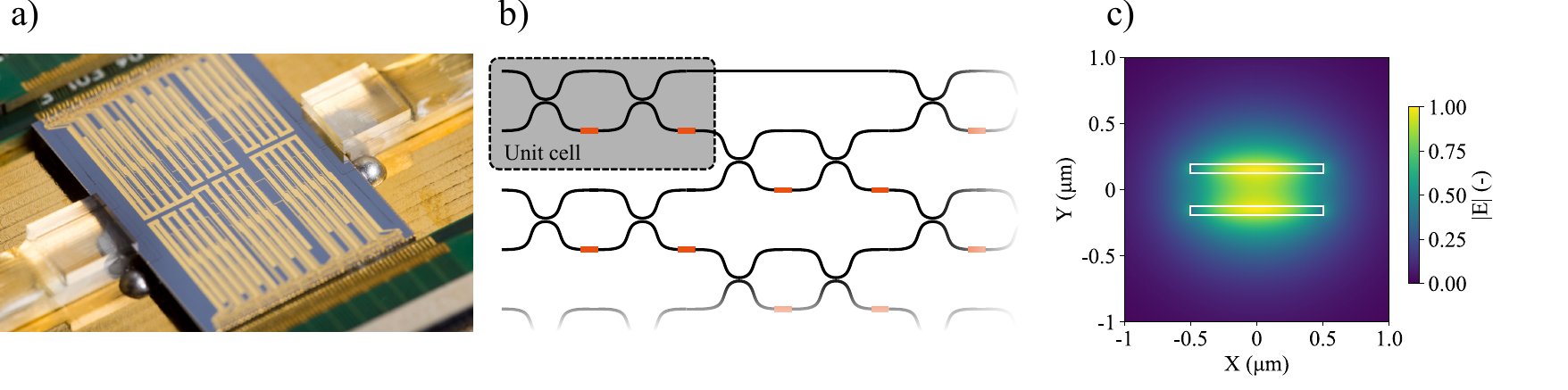}
  \caption{\textbf{a)} Photograph of the 12-mode photonic processor chip (\SI{16}{mm} $\times$ \SI{22}{mm}). \textbf{b)} Functional design of the photonic network consisting of interconnected waveguide paths (black) and thermo-optic phase actuators (red). The network is built by repeating the unit cell, which contains a tunable beam splitter (TBS) and external phase shifter (PS). \textbf{c)} Simulated mode field profile of the supported TE mode for \SI{940}{nm} of the optimal waveguide design. The simulation was performed by a fully vectorial 2D eigenmode calculation using Lumerical MODE solutions. The white lines indicate the symmetric double-stripe SiN waveguide within the silica glass cladding.}
  \label{fig:chip}
\end{figure}

The quantum photonic processor consists out of three parts: a photonic integrated circuit of stoichiometric SiN symmetric double-stripe waveguides with the TriPleX technology \cite{c_g_h_roeloffzen_low-loss_2018}, control electronics and control software. The device is optically packaged with input and output fiber arrays and electronically connected by wire-bonds to a control PCB, as shown in Fig.\ref{fig:chip}a. The photonic chip contains a mesh of interconnected waveguides with 12 input and 12 output modes, enabling the implementation of arbitrary $12 \times 12$ unitary matrix transformations on the input light states. The network consists of a repeating unit cell that contains a tunable Mach-Zehnder interferometer (MZI) to set the relative amplitudes and an external phase shifter to set the relative phase of the two waveguides, as shown in Fig.\ref{fig:chip}b, together forming an arbitrarily tunable beam-splitter. Both tunable elements are realized with thermo-optic phase actuators, of which there are 132 distributed over 66 unit cells. The phases required for a specific matrix transformation are determined based on the method of \cite{clements_optimal_2016}. The phase actuators implement phase shifts by resistive heating, locally elevating the waveguide temperature and thus inducing a change in refractive index. The control electronics and control software ensure a stable device temperature during operation and implement the corresponding phases for arbitrary $12 \times 12$ unitary matrices.

The design of the SiN photonic chip was performed for operation around a central wavelength of \SI{940}{nm} to allow for compatibility with the InGaAs emission wavelength range. The geometry consists of two identical SiN stripes separated by a gap and embedded in a silica glass cladding, as shown in Fig.\ref{fig:chip}c. The waveguide geometry was constrained to only support a single (TE) mode, which also has low bend losses, allowing for small bend radii. We found an optimal design with a stripe width of \SI{1000}{nm}, stripe height of \SI{65}{nm} and a vertical gap of \SI{250}{nm}. This design has low bend losses down to a radius of \SI{110}{\micro\meter}. Finally, we designed a horizontal taper for a fiber-to-chip coupling spot-size converter with an expected coupling loss of \SI{0.4}{dB}, which is crucial for quantum photonic computing since photon loss should be minimal to reduce overhead.

\section{\label{sec:setup} Experimental Results}
The quantum photonic processor was characterized by injecting coherent light at a wavelength of \SI{940}{nm} into the device (TOPTICA FF Pro QD tunable laser). The light was guided into the photonic processor using a $1\times12$ PM fiber switch (Fibermart), followed by detection at an array of photodiodes (Thorlabs FGA01FC). 

\begin{figure*}
\includegraphics{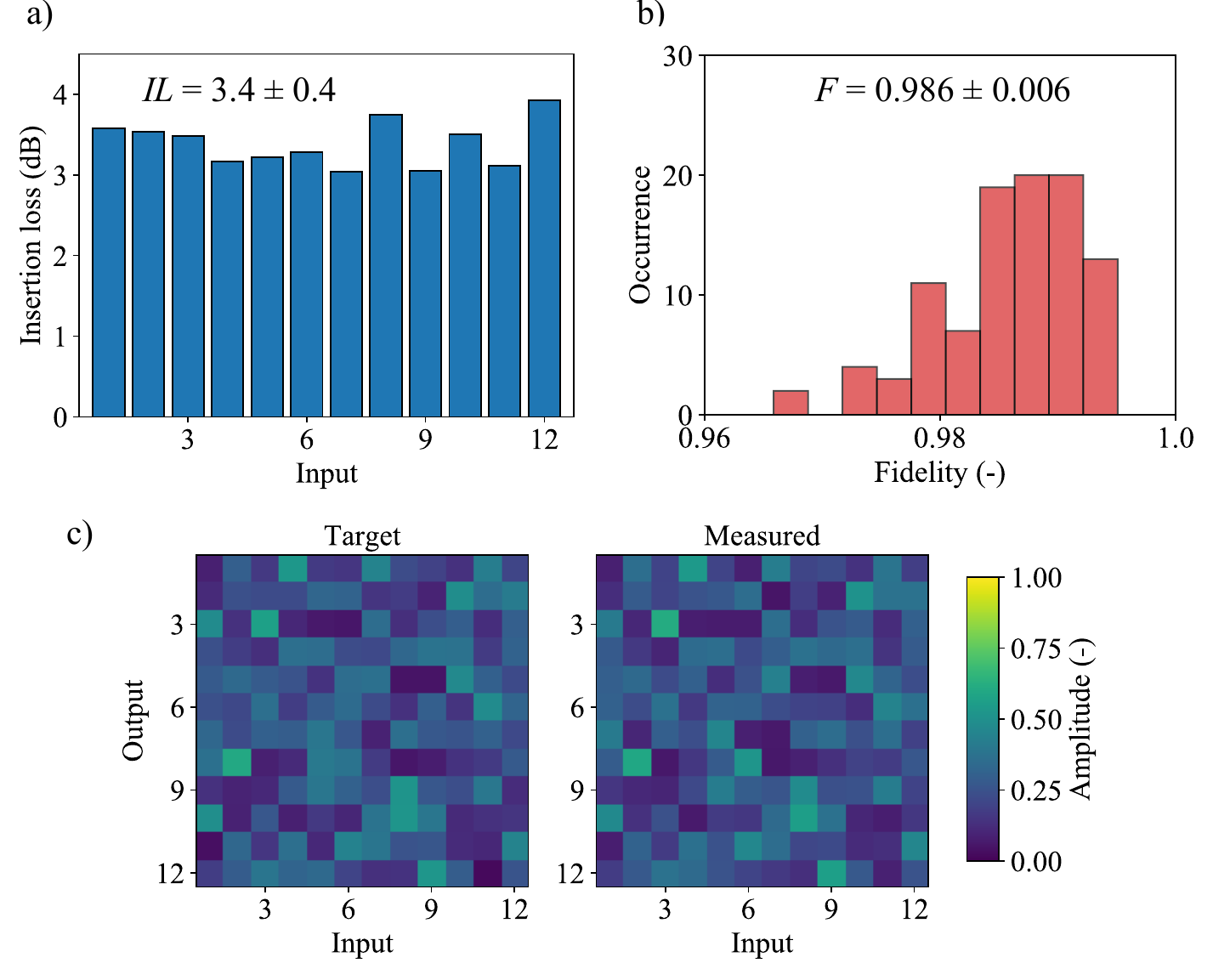}
  \caption{Summary of the characterization. \textbf{a)} Measured insertion loss for each mode on the photonic chip. \textbf{b)} Distribution of the measured amplitude fidelities for 100 Haar-random matrices. \textbf{c)} Comparison between the amplitude distribution of an ideal Haar-random unitary matrix (target) with its experimental implementation (data).}
  \label{fig:results}
\end{figure*}

First, the phase-voltage relationship of the tunable thermo-optic phase actuators was characterized. Voltages were applied to each phase actuator while recording the output powers at the photodiodes, from which the phase response of each element was obtained. The average power for a $2\pi$ phase shift was \SI{310}{mW}. Furthermore, we find that our directional couplers operate very close to the \SI{3}{dB} power coupling point at the wavelength of \SI{940}{nm}, as evidenced by an average tunable MZI extinction ratio of \SI{22.4}{dB}. Next, the average insertion loss of the photonic processor was measured to be \SI{3.4}{dB}, as shown in Fig.\,\ref{fig:results}a. This includes the loss of the fiber-to-chip and chip-to-fiber coupling and propagation through the network with a path length of \SI{10.7}{cm}. Finally, the reconfigurability of the photonic processor was characterized by generating, implementing, and measuring 100 Haar-random matrices on the network. The output intensity distribution was measured for each input mode of the matrices, followed by normalizing the data by correcting for the port-dependent coupling losses and photodiode dark counts. The fidelity measure of the matrices was then determined by $F = \Tr(|U^+|\cdot|U_{\rm exp}|)/12$, where $U$ is the target matrix and $U_{\rm exp}$ the measured, normalized matrix. We obtained an average fidelity of $F=0.986\pm0.006$, as shown in Fig.\,\ref{fig:results}b. As an example, Fig.\,\ref{fig:results}c shows a comparison between one of the target Haar-random matrices (left), and the measured results from the implementation of that matrix on the processor with a fidelity of $F=0.987$ (right). 

\section{\label{sec:discussion_conclusions}Discussion and Conclusions}
Our work demonstrates the first large-scale, universally reconfigurable quantum photonic processor that operates at a wavelength range outside of the prevalent telecom C-band. The device has low insertion loss and a high degree of reconfigurability. Despite its smaller size and the reduced optical path length per mode, the insertion losses are higher than the previously demonstrated value of \SI{2.9}{dB/mode} for a SiN 20-mode quantum photonic processor \cite{Taballione_20mode_2022}. The higher loss results from the increased Rayleigh scattering at the lower wavelength range than that of the C band, which can be mitigated by improving the fabrication process. Furthermore, this device has more efficient heaters that require a lower power for a $2\pi$ phase shift than the other demonstrated large-scale SiN quantum photonic processors \cite{taballione_universal_2021, Taballione_20mode_2022}. This is beneficial for a lower thermal load, allowing a higher density of heaters integrated on the photonic circuit, and for a more precise reconfigurability and a higher fidelity. In fact, this improved the matrix fidelity compared with the 12-mode quantum photonic processor of \cite{taballione_universal_2021}.

The used material platform SiN benefits from a wide transparency window enabling operation at a wide range of wavelengths. The demonstrated operating wavelength of \SI{940}{nm} is of specific interest for InGaAs-based quantum dot single photon sources that benefit from scalable fabrication and integration techniques. Furthermore, SiN is a highly versatile platform enabling fast switching \cite{epping_ultra-low-power_2017}, lowest propagation losses \cite{bauters_planar_2011} and superconducting single-photon detectors \cite{schuck_nbtin_2013,schuck_quantum_2016}, thus presenting itself as a viable path towards fully integrated quantum computers.

\bibliography{referencesOLD}

\begin{thebibliography}{31}%
\makeatletter
\providecommand \@ifxundefined [1]{%
 \@ifx{#1\undefined}
}%
\providecommand \@ifnum [1]{%
 \ifnum #1\expandafter \@firstoftwo
 \else \expandafter \@secondoftwo
 \fi
}%
\providecommand \@ifx [1]{%
 \ifx #1\expandafter \@firstoftwo
 \else \expandafter \@secondoftwo
 \fi
}%
\providecommand \natexlab [1]{#1}%
\providecommand \enquote  [1]{``#1''}%
\providecommand \bibnamefont  [1]{#1}%
\providecommand \bibfnamefont [1]{#1}%
\providecommand \citenamefont [1]{#1}%
\providecommand \href@noop [0]{\@secondoftwo}%
\providecommand \href [0]{\begingroup \@sanitize@url \@href}%
\providecommand \@href[1]{\@@startlink{#1}\@@href}%
\providecommand \@@href[1]{\endgroup#1\@@endlink}%
\providecommand \@sanitize@url [0]{\catcode `\\12\catcode `\$12\catcode
  `\&12\catcode `\#12\catcode `\^12\catcode `\_12\catcode `\%12\relax}%
\providecommand \@@startlink[1]{}%
\providecommand \@@endlink[0]{}%
\providecommand \url  [0]{\begingroup\@sanitize@url \@url }%
\providecommand \@url [1]{\endgroup\@href {#1}{\urlprefix }}%
\providecommand \urlprefix  [0]{URL }%
\providecommand \Eprint [0]{\href }%
\providecommand \doibase [0]{https://doi.org/}%
\providecommand \selectlanguage [0]{\@gobble}%
\providecommand \bibinfo  [0]{\@secondoftwo}%
\providecommand \bibfield  [0]{\@secondoftwo}%
\providecommand \translation [1]{[#1]}%
\providecommand \BibitemOpen [0]{}%
\providecommand \bibitemStop [0]{}%
\providecommand \bibitemNoStop [0]{.\EOS\space}%
\providecommand \EOS [0]{\spacefactor3000\relax}%
\providecommand \BibitemShut  [1]{\csname bibitem#1\endcsname}%
\let\auto@bib@innerbib\@empty
%</preamble>
\bibitem [{\citenamefont {Zhong}\ \emph {et~al.}(2020)\citenamefont {Zhong},
  \citenamefont {Wang}, \citenamefont {Deng}, \citenamefont {Chen},
  \citenamefont {Peng}, \citenamefont {Luo}, \citenamefont {Qin}, \citenamefont
  {Wu}, \citenamefont {Ding}, \citenamefont {Hu}, \citenamefont {Hu},
  \citenamefont {Yang}, \citenamefont {Zhang}, \citenamefont {Li},
  \citenamefont {Li}, \citenamefont {Jiang}, \citenamefont {Gan}, \citenamefont
  {Yang}, \citenamefont {You}, \citenamefont {Wang}, \citenamefont {Li},
  \citenamefont {Liu}, \citenamefont {Lu},\ and\ \citenamefont
  {Pan}}]{zhong_quantum_2020}%
  \BibitemOpen
  \bibfield  {author} {\bibinfo {author} {\bibfnamefont {H.-S.}\ \bibnamefont
  {Zhong}}, \bibinfo {author} {\bibfnamefont {H.}~\bibnamefont {Wang}},
  \bibinfo {author} {\bibfnamefont {Y.-H.}\ \bibnamefont {Deng}}, \bibinfo
  {author} {\bibfnamefont {M.-C.}\ \bibnamefont {Chen}}, \bibinfo {author}
  {\bibfnamefont {L.-C.}\ \bibnamefont {Peng}}, \bibinfo {author}
  {\bibfnamefont {Y.-H.}\ \bibnamefont {Luo}}, \bibinfo {author} {\bibfnamefont
  {J.}~\bibnamefont {Qin}}, \bibinfo {author} {\bibfnamefont {D.}~\bibnamefont
  {Wu}}, \bibinfo {author} {\bibfnamefont {X.}~\bibnamefont {Ding}}, \bibinfo
  {author} {\bibfnamefont {Y.}~\bibnamefont {Hu}}, \bibinfo {author}
  {\bibfnamefont {P.}~\bibnamefont {Hu}}, \bibinfo {author} {\bibfnamefont
  {X.-Y.}\ \bibnamefont {Yang}}, \bibinfo {author} {\bibfnamefont {W.-J.}\
  \bibnamefont {Zhang}}, \bibinfo {author} {\bibfnamefont {H.}~\bibnamefont
  {Li}}, \bibinfo {author} {\bibfnamefont {Y.}~\bibnamefont {Li}}, \bibinfo
  {author} {\bibfnamefont {X.}~\bibnamefont {Jiang}}, \bibinfo {author}
  {\bibfnamefont {L.}~\bibnamefont {Gan}}, \bibinfo {author} {\bibfnamefont
  {G.}~\bibnamefont {Yang}}, \bibinfo {author} {\bibfnamefont {L.}~\bibnamefont
  {You}}, \bibinfo {author} {\bibfnamefont {Z.}~\bibnamefont {Wang}}, \bibinfo
  {author} {\bibfnamefont {L.}~\bibnamefont {Li}}, \bibinfo {author}
  {\bibfnamefont {N.-L.}\ \bibnamefont {Liu}}, \bibinfo {author} {\bibfnamefont
  {C.-Y.}\ \bibnamefont {Lu}},\ and\ \bibinfo {author} {\bibfnamefont {J.-W.}\
  \bibnamefont {Pan}},\ }\bibfield  {title} {\bibinfo {title} {Quantum
  computational advantage using photons},\ }\href
  {https://doi.org/10.1126/science.abe8770} {\bibfield  {journal} {\bibinfo
  {journal} {Science}\ }\textbf {\bibinfo {volume} {370}},\ \bibinfo {pages}
  {1460} (\bibinfo {year} {2020})},\ \bibinfo {note} {\_eprint:
  https://www.science.org/doi/pdf/10.1126/science.abe8770}\BibitemShut
  {NoStop}%
\bibitem [{\citenamefont {Arute}\ \emph {et~al.}(2019)\citenamefont {Arute},
  \citenamefont {Arya}, \citenamefont {Babbush}, \citenamefont {Bacon},
  \citenamefont {Bardin}, \citenamefont {Barends}, \citenamefont {Biswas},
  \citenamefont {Boixo}, \citenamefont {Brandao}, \citenamefont {Buell},
  \citenamefont {Burkett}, \citenamefont {Chen}, \citenamefont {Chen},
  \citenamefont {Chiaro}, \citenamefont {Collins}, \citenamefont {Courtney},
  \citenamefont {Dunsworth}, \citenamefont {Farhi}, \citenamefont {Foxen},
  \citenamefont {Fowler}, \citenamefont {Gidney}, \citenamefont {Giustina},
  \citenamefont {Graff}, \citenamefont {Guerin}, \citenamefont {Habegger},
  \citenamefont {Harrigan}, \citenamefont {Hartmann}, \citenamefont {Ho},
  \citenamefont {Hoffmann}, \citenamefont {Huang}, \citenamefont {Humble},
  \citenamefont {Isakov}, \citenamefont {Jeffrey}, \citenamefont {Jiang},
  \citenamefont {Kafri}, \citenamefont {Kechedzhi}, \citenamefont {Kelly},
  \citenamefont {Klimov}, \citenamefont {Knysh}, \citenamefont {Korotkov},
  \citenamefont {Kostritsa}, \citenamefont {Landhuis}, \citenamefont
  {Lindmark}, \citenamefont {Lucero}, \citenamefont {Lyakh}, \citenamefont
  {Mandr{\`a}}, \citenamefont {McClean}, \citenamefont {McEwen}, \citenamefont
  {Megrant}, \citenamefont {Mi}, \citenamefont {Michielsen}, \citenamefont
  {Mohseni}, \citenamefont {Mutus}, \citenamefont {Naaman}, \citenamefont
  {Neeley}, \citenamefont {Neill}, \citenamefont {Niu}, \citenamefont {Ostby},
  \citenamefont {Petukhov}, \citenamefont {Platt}, \citenamefont {Quintana},
  \citenamefont {Rieffel}, \citenamefont {Roushan}, \citenamefont {Rubin},
  \citenamefont {Sank}, \citenamefont {Satzinger}, \citenamefont {Smelyanskiy},
  \citenamefont {Sung}, \citenamefont {Trevithick}, \citenamefont
  {Vainsencher}, \citenamefont {Villalonga}, \citenamefont {White},
  \citenamefont {Yao}, \citenamefont {Yeh}, \citenamefont {Zalcman},
  \citenamefont {Neven},\ and\ \citenamefont {Martinis}}]{Arute_Google_2019}%
  \BibitemOpen
  \bibfield  {author} {\bibinfo {author} {\bibfnamefont {F.}~\bibnamefont
  {Arute}}, \bibinfo {author} {\bibfnamefont {K.}~\bibnamefont {Arya}},
  \bibinfo {author} {\bibfnamefont {R.}~\bibnamefont {Babbush}}, \bibinfo
  {author} {\bibfnamefont {D.}~\bibnamefont {Bacon}}, \bibinfo {author}
  {\bibfnamefont {J.~C.}\ \bibnamefont {Bardin}}, \bibinfo {author}
  {\bibfnamefont {R.}~\bibnamefont {Barends}}, \bibinfo {author} {\bibfnamefont
  {R.}~\bibnamefont {Biswas}}, \bibinfo {author} {\bibfnamefont
  {S.}~\bibnamefont {Boixo}}, \bibinfo {author} {\bibfnamefont {F.~G. S.~L.}\
  \bibnamefont {Brandao}}, \bibinfo {author} {\bibfnamefont {D.~A.}\
  \bibnamefont {Buell}}, \bibinfo {author} {\bibfnamefont {B.}~\bibnamefont
  {Burkett}}, \bibinfo {author} {\bibfnamefont {Y.}~\bibnamefont {Chen}},
  \bibinfo {author} {\bibfnamefont {Z.}~\bibnamefont {Chen}}, \bibinfo {author}
  {\bibfnamefont {B.}~\bibnamefont {Chiaro}}, \bibinfo {author} {\bibfnamefont
  {R.}~\bibnamefont {Collins}}, \bibinfo {author} {\bibfnamefont
  {W.}~\bibnamefont {Courtney}}, \bibinfo {author} {\bibfnamefont
  {A.}~\bibnamefont {Dunsworth}}, \bibinfo {author} {\bibfnamefont
  {E.}~\bibnamefont {Farhi}}, \bibinfo {author} {\bibfnamefont
  {B.}~\bibnamefont {Foxen}}, \bibinfo {author} {\bibfnamefont
  {A.}~\bibnamefont {Fowler}}, \bibinfo {author} {\bibfnamefont
  {C.}~\bibnamefont {Gidney}}, \bibinfo {author} {\bibfnamefont
  {M.}~\bibnamefont {Giustina}}, \bibinfo {author} {\bibfnamefont
  {R.}~\bibnamefont {Graff}}, \bibinfo {author} {\bibfnamefont
  {K.}~\bibnamefont {Guerin}}, \bibinfo {author} {\bibfnamefont
  {S.}~\bibnamefont {Habegger}}, \bibinfo {author} {\bibfnamefont {M.~P.}\
  \bibnamefont {Harrigan}}, \bibinfo {author} {\bibfnamefont {M.~J.}\
  \bibnamefont {Hartmann}}, \bibinfo {author} {\bibfnamefont {A.}~\bibnamefont
  {Ho}}, \bibinfo {author} {\bibfnamefont {M.}~\bibnamefont {Hoffmann}},
  \bibinfo {author} {\bibfnamefont {T.}~\bibnamefont {Huang}}, \bibinfo
  {author} {\bibfnamefont {T.~S.}\ \bibnamefont {Humble}}, \bibinfo {author}
  {\bibfnamefont {S.~V.}\ \bibnamefont {Isakov}}, \bibinfo {author}
  {\bibfnamefont {E.}~\bibnamefont {Jeffrey}}, \bibinfo {author} {\bibfnamefont
  {Z.}~\bibnamefont {Jiang}}, \bibinfo {author} {\bibfnamefont
  {D.}~\bibnamefont {Kafri}}, \bibinfo {author} {\bibfnamefont
  {K.}~\bibnamefont {Kechedzhi}}, \bibinfo {author} {\bibfnamefont
  {J.}~\bibnamefont {Kelly}}, \bibinfo {author} {\bibfnamefont {P.~V.}\
  \bibnamefont {Klimov}}, \bibinfo {author} {\bibfnamefont {S.}~\bibnamefont
  {Knysh}}, \bibinfo {author} {\bibfnamefont {A.}~\bibnamefont {Korotkov}},
  \bibinfo {author} {\bibfnamefont {F.}~\bibnamefont {Kostritsa}}, \bibinfo
  {author} {\bibfnamefont {D.}~\bibnamefont {Landhuis}}, \bibinfo {author}
  {\bibfnamefont {M.}~\bibnamefont {Lindmark}}, \bibinfo {author}
  {\bibfnamefont {E.}~\bibnamefont {Lucero}}, \bibinfo {author} {\bibfnamefont
  {D.}~\bibnamefont {Lyakh}}, \bibinfo {author} {\bibfnamefont
  {S.}~\bibnamefont {Mandr{\`a}}}, \bibinfo {author} {\bibfnamefont {J.~R.}\
  \bibnamefont {McClean}}, \bibinfo {author} {\bibfnamefont {M.}~\bibnamefont
  {McEwen}}, \bibinfo {author} {\bibfnamefont {A.}~\bibnamefont {Megrant}},
  \bibinfo {author} {\bibfnamefont {X.}~\bibnamefont {Mi}}, \bibinfo {author}
  {\bibfnamefont {K.}~\bibnamefont {Michielsen}}, \bibinfo {author}
  {\bibfnamefont {M.}~\bibnamefont {Mohseni}}, \bibinfo {author} {\bibfnamefont
  {J.}~\bibnamefont {Mutus}}, \bibinfo {author} {\bibfnamefont
  {O.}~\bibnamefont {Naaman}}, \bibinfo {author} {\bibfnamefont
  {M.}~\bibnamefont {Neeley}}, \bibinfo {author} {\bibfnamefont
  {C.}~\bibnamefont {Neill}}, \bibinfo {author} {\bibfnamefont {M.~Y.}\
  \bibnamefont {Niu}}, \bibinfo {author} {\bibfnamefont {E.}~\bibnamefont
  {Ostby}}, \bibinfo {author} {\bibfnamefont {A.}~\bibnamefont {Petukhov}},
  \bibinfo {author} {\bibfnamefont {J.~C.}\ \bibnamefont {Platt}}, \bibinfo
  {author} {\bibfnamefont {C.}~\bibnamefont {Quintana}}, \bibinfo {author}
  {\bibfnamefont {E.~G.}\ \bibnamefont {Rieffel}}, \bibinfo {author}
  {\bibfnamefont {P.}~\bibnamefont {Roushan}}, \bibinfo {author} {\bibfnamefont
  {N.~C.}\ \bibnamefont {Rubin}}, \bibinfo {author} {\bibfnamefont
  {D.}~\bibnamefont {Sank}}, \bibinfo {author} {\bibfnamefont {K.~J.}\
  \bibnamefont {Satzinger}}, \bibinfo {author} {\bibfnamefont {V.}~\bibnamefont
  {Smelyanskiy}}, \bibinfo {author} {\bibfnamefont {K.~J.}\ \bibnamefont
  {Sung}}, \bibinfo {author} {\bibfnamefont {M.~D.}\ \bibnamefont
  {Trevithick}}, \bibinfo {author} {\bibfnamefont {A.}~\bibnamefont
  {Vainsencher}}, \bibinfo {author} {\bibfnamefont {B.}~\bibnamefont
  {Villalonga}}, \bibinfo {author} {\bibfnamefont {T.}~\bibnamefont {White}},
  \bibinfo {author} {\bibfnamefont {Z.~J.}\ \bibnamefont {Yao}}, \bibinfo
  {author} {\bibfnamefont {P.}~\bibnamefont {Yeh}}, \bibinfo {author}
  {\bibfnamefont {A.}~\bibnamefont {Zalcman}}, \bibinfo {author} {\bibfnamefont
  {H.}~\bibnamefont {Neven}},\ and\ \bibinfo {author} {\bibfnamefont {J.~M.}\
  \bibnamefont {Martinis}},\ }\bibfield  {title} {\bibinfo {title} {Quantum
  supremacy using a programmable superconducting processor},\ }\href
  {https://doi.org/10.1038/s41586-019-1666-5} {\bibfield  {journal} {\bibinfo
  {journal} {Nature}\ }\textbf {\bibinfo {volume} {574}},\ \bibinfo {pages}
  {505} (\bibinfo {year} {2019})}\BibitemShut {NoStop}%
\bibitem [{\citenamefont {Wang}\ \emph {et~al.}(2020)\citenamefont {Wang},
  \citenamefont {Sciarrino}, \citenamefont {Laing},\ and\ \citenamefont
  {Thompson}}]{Wang_processor_2020}%
  \BibitemOpen
  \bibfield  {author} {\bibinfo {author} {\bibfnamefont {J.}~\bibnamefont
  {Wang}}, \bibinfo {author} {\bibfnamefont {F.}~\bibnamefont {Sciarrino}},
  \bibinfo {author} {\bibfnamefont {A.}~\bibnamefont {Laing}},\ and\ \bibinfo
  {author} {\bibfnamefont {M.~G.}\ \bibnamefont {Thompson}},\ }\bibfield
  {title} {\bibinfo {title} {Integrated photonic quantum technologies},\ }\href
  {https://doi.org/10.1038/s41566-019-0532-1} {\bibfield  {journal} {\bibinfo
  {journal} {Nature Photonics}\ }\textbf {\bibinfo {volume} {14}},\ \bibinfo
  {pages} {273} (\bibinfo {year} {2020})}\BibitemShut {NoStop}%
\bibitem [{\citenamefont {Somaschi}\ \emph {et~al.}(2016)\citenamefont
  {Somaschi}, \citenamefont {Giesz}, \citenamefont {De~Santis}, \citenamefont
  {Loredo}, \citenamefont {Almeida}, \citenamefont {Hornecker}, \citenamefont
  {Portalupi}, \citenamefont {Grange}, \citenamefont {Ant{\'o}n}, \citenamefont
  {Demory}, \citenamefont {G{\'o}mez}, \citenamefont {Sagnes}, \citenamefont
  {Lanzillotti-Kimura}, \citenamefont {Lema{\'i}tre}, \citenamefont {Auffeves},
  \citenamefont {White}, \citenamefont {Lanco},\ and\ \citenamefont
  {Senellart}}]{Somaschi_QD_2016}%
  \BibitemOpen
  \bibfield  {author} {\bibinfo {author} {\bibfnamefont {N.}~\bibnamefont
  {Somaschi}}, \bibinfo {author} {\bibfnamefont {V.}~\bibnamefont {Giesz}},
  \bibinfo {author} {\bibfnamefont {L.}~\bibnamefont {De~Santis}}, \bibinfo
  {author} {\bibfnamefont {J.~C.}\ \bibnamefont {Loredo}}, \bibinfo {author}
  {\bibfnamefont {M.~P.}\ \bibnamefont {Almeida}}, \bibinfo {author}
  {\bibfnamefont {G.}~\bibnamefont {Hornecker}}, \bibinfo {author}
  {\bibfnamefont {S.~L.}\ \bibnamefont {Portalupi}}, \bibinfo {author}
  {\bibfnamefont {T.}~\bibnamefont {Grange}}, \bibinfo {author} {\bibfnamefont
  {C.}~\bibnamefont {Ant{\'o}n}}, \bibinfo {author} {\bibfnamefont
  {J.}~\bibnamefont {Demory}}, \bibinfo {author} {\bibfnamefont
  {C.}~\bibnamefont {G{\'o}mez}}, \bibinfo {author} {\bibfnamefont
  {I.}~\bibnamefont {Sagnes}}, \bibinfo {author} {\bibfnamefont {N.~D.}\
  \bibnamefont {Lanzillotti-Kimura}}, \bibinfo {author} {\bibfnamefont
  {A.}~\bibnamefont {Lema{\'i}tre}}, \bibinfo {author} {\bibfnamefont
  {A.}~\bibnamefont {Auffeves}}, \bibinfo {author} {\bibfnamefont {A.~G.}\
  \bibnamefont {White}}, \bibinfo {author} {\bibfnamefont {L.}~\bibnamefont
  {Lanco}},\ and\ \bibinfo {author} {\bibfnamefont {P.}~\bibnamefont
  {Senellart}},\ }\bibfield  {title} {\bibinfo {title} {Near-optimal
  single-photon sources in the solid state},\ }\href
  {https://doi.org/10.1038/nphoton.2016.23} {\bibfield  {journal} {\bibinfo
  {journal} {Nature Photonics}\ }\textbf {\bibinfo {volume} {10}},\ \bibinfo
  {pages} {340} (\bibinfo {year} {2016})}\BibitemShut {NoStop}%
\bibitem [{\citenamefont {Ding}\ \emph {et~al.}(2016)\citenamefont {Ding},
  \citenamefont {He}, \citenamefont {Duan}, \citenamefont {Gregersen},
  \citenamefont {Chen}, \citenamefont {Unsleber}, \citenamefont {Maier},
  \citenamefont {Schneider}, \citenamefont {Kamp}, \citenamefont {H\"ofling},
  \citenamefont {Lu},\ and\ \citenamefont {Pan}}]{Xing_QD_2016}%
  \BibitemOpen
  \bibfield  {author} {\bibinfo {author} {\bibfnamefont {X.}~\bibnamefont
  {Ding}}, \bibinfo {author} {\bibfnamefont {Y.}~\bibnamefont {He}}, \bibinfo
  {author} {\bibfnamefont {Z.-C.}\ \bibnamefont {Duan}}, \bibinfo {author}
  {\bibfnamefont {N.}~\bibnamefont {Gregersen}}, \bibinfo {author}
  {\bibfnamefont {M.-C.}\ \bibnamefont {Chen}}, \bibinfo {author}
  {\bibfnamefont {S.}~\bibnamefont {Unsleber}}, \bibinfo {author}
  {\bibfnamefont {S.}~\bibnamefont {Maier}}, \bibinfo {author} {\bibfnamefont
  {C.}~\bibnamefont {Schneider}}, \bibinfo {author} {\bibfnamefont
  {M.}~\bibnamefont {Kamp}}, \bibinfo {author} {\bibfnamefont {S.}~\bibnamefont
  {H\"ofling}}, \bibinfo {author} {\bibfnamefont {C.-Y.}\ \bibnamefont {Lu}},\
  and\ \bibinfo {author} {\bibfnamefont {J.-W.}\ \bibnamefont {Pan}},\
  }\bibfield  {title} {\bibinfo {title} {On-demand single photons with high
  extraction efficiency and near-unity indistinguishability from a resonantly
  driven quantum dot in a micropillar},\ }\href
  {https://doi.org/10.1103/PhysRevLett.116.020401} {\bibfield  {journal}
  {\bibinfo  {journal} {Phys. Rev. Lett.}\ }\textbf {\bibinfo {volume} {116}},\
  \bibinfo {pages} {020401} (\bibinfo {year} {2016})}\BibitemShut {NoStop}%
\bibitem [{\citenamefont {Ollivier}\ \emph {et~al.}(2020)\citenamefont
  {Ollivier}, \citenamefont {Maillette~de Buy~Wenniger}, \citenamefont
  {Thomas}, \citenamefont {Wein}, \citenamefont {Harouri}, \citenamefont
  {Coppola}, \citenamefont {Hilaire}, \citenamefont {Millet}, \citenamefont
  {Lema{\^i}tre}, \citenamefont {Sagnes}, \citenamefont {Krebs}, \citenamefont
  {Lanco}, \citenamefont {Loredo}, \citenamefont {Ant{\'o}n}, \citenamefont
  {Somaschi},\ and\ \citenamefont {Senellart}}]{Ollivier_QD_2020}%
  \BibitemOpen
  \bibfield  {author} {\bibinfo {author} {\bibfnamefont {H.}~\bibnamefont
  {Ollivier}}, \bibinfo {author} {\bibfnamefont {I.}~\bibnamefont {Maillette~de
  Buy~Wenniger}}, \bibinfo {author} {\bibfnamefont {S.}~\bibnamefont {Thomas}},
  \bibinfo {author} {\bibfnamefont {S.~C.}\ \bibnamefont {Wein}}, \bibinfo
  {author} {\bibfnamefont {A.}~\bibnamefont {Harouri}}, \bibinfo {author}
  {\bibfnamefont {G.}~\bibnamefont {Coppola}}, \bibinfo {author} {\bibfnamefont
  {P.}~\bibnamefont {Hilaire}}, \bibinfo {author} {\bibfnamefont
  {C.}~\bibnamefont {Millet}}, \bibinfo {author} {\bibfnamefont
  {A.}~\bibnamefont {Lema{\^i}tre}}, \bibinfo {author} {\bibfnamefont
  {I.}~\bibnamefont {Sagnes}}, \bibinfo {author} {\bibfnamefont
  {O.}~\bibnamefont {Krebs}}, \bibinfo {author} {\bibfnamefont
  {L.}~\bibnamefont {Lanco}}, \bibinfo {author} {\bibfnamefont {J.~C.}\
  \bibnamefont {Loredo}}, \bibinfo {author} {\bibfnamefont {C.}~\bibnamefont
  {Ant{\'o}n}}, \bibinfo {author} {\bibfnamefont {N.}~\bibnamefont
  {Somaschi}},\ and\ \bibinfo {author} {\bibfnamefont {P.}~\bibnamefont
  {Senellart}},\ }\bibfield  {title} {\bibinfo {title} {Reproducibility of
  high-performance quantum dot single-photon sources},\ }\href
  {https://doi.org/10.1021/acsphotonics.9b01805} {\bibfield  {journal}
  {\bibinfo  {journal} {ACS Photonics}\ }\textbf {\bibinfo {volume} {7}},\
  \bibinfo {pages} {1050} (\bibinfo {year} {2020})}\BibitemShut {NoStop}%
\bibitem [{\citenamefont {Uppu}\ \emph {et~al.}(2020)\citenamefont {Uppu},
  \citenamefont {Pedersen}, \citenamefont {Wang}, \citenamefont {Olesen},
  \citenamefont {Papon}, \citenamefont {Zhou}, \citenamefont {Midolo},
  \citenamefont {Scholz}, \citenamefont {Wieck}, \citenamefont {Ludwig},\ and\
  \citenamefont {Lodahl}}]{Uppu_QD_2020}%
  \BibitemOpen
  \bibfield  {author} {\bibinfo {author} {\bibfnamefont {R.}~\bibnamefont
  {Uppu}}, \bibinfo {author} {\bibfnamefont {F.~T.}\ \bibnamefont {Pedersen}},
  \bibinfo {author} {\bibfnamefont {Y.}~\bibnamefont {Wang}}, \bibinfo {author}
  {\bibfnamefont {C.~T.}\ \bibnamefont {Olesen}}, \bibinfo {author}
  {\bibfnamefont {C.}~\bibnamefont {Papon}}, \bibinfo {author} {\bibfnamefont
  {X.}~\bibnamefont {Zhou}}, \bibinfo {author} {\bibfnamefont {L.}~\bibnamefont
  {Midolo}}, \bibinfo {author} {\bibfnamefont {S.}~\bibnamefont {Scholz}},
  \bibinfo {author} {\bibfnamefont {A.~D.}\ \bibnamefont {Wieck}}, \bibinfo
  {author} {\bibfnamefont {A.}~\bibnamefont {Ludwig}},\ and\ \bibinfo {author}
  {\bibfnamefont {P.}~\bibnamefont {Lodahl}},\ }\bibfield  {title} {\bibinfo
  {title} {Scalable integrated single-photon source},\ }\href
  {https://doi.org/10.1126/sciadv.abc8268} {\bibfield  {journal} {\bibinfo
  {journal} {Science Advances}\ }\textbf {\bibinfo {volume} {6}},\ \bibinfo
  {pages} {eabc8268} (\bibinfo {year} {2020})},\ \Eprint
  {https://arxiv.org/abs/https://www.science.org/doi/pdf/10.1126/sciadv.abc8268}
  {https://www.science.org/doi/pdf/10.1126/sciadv.abc8268} \BibitemShut
  {NoStop}%
\bibitem [{\citenamefont {Senellart}\ \emph {et~al.}(2017)\citenamefont
  {Senellart}, \citenamefont {Solomon},\ and\ \citenamefont
  {White}}]{Senellart_QD_2017}%
  \BibitemOpen
  \bibfield  {author} {\bibinfo {author} {\bibfnamefont {P.}~\bibnamefont
  {Senellart}}, \bibinfo {author} {\bibfnamefont {G.}~\bibnamefont {Solomon}},\
  and\ \bibinfo {author} {\bibfnamefont {A.}~\bibnamefont {White}},\ }\bibfield
   {title} {\bibinfo {title} {High-performance semiconductor quantum-dot
  single-photon sources},\ }\href {https://doi.org/10.1038/nnano.2017.218}
  {\bibfield  {journal} {\bibinfo  {journal} {Nature Nanotechnology}\ }\textbf
  {\bibinfo {volume} {12}},\ \bibinfo {pages} {1026} (\bibinfo {year}
  {2017})}\BibitemShut {NoStop}%
\bibitem [{\citenamefont {Loredo}\ \emph {et~al.}(2017)\citenamefont {Loredo},
  \citenamefont {Broome}, \citenamefont {Hilaire}, \citenamefont {Gazzano},
  \citenamefont {Sagnes}, \citenamefont {Lemaitre}, \citenamefont {Almeida},
  \citenamefont {Senellart},\ and\ \citenamefont {White}}]{Loredo_BS_2017}%
  \BibitemOpen
  \bibfield  {author} {\bibinfo {author} {\bibfnamefont {J.~C.}\ \bibnamefont
  {Loredo}}, \bibinfo {author} {\bibfnamefont {M.~A.}\ \bibnamefont {Broome}},
  \bibinfo {author} {\bibfnamefont {P.}~\bibnamefont {Hilaire}}, \bibinfo
  {author} {\bibfnamefont {O.}~\bibnamefont {Gazzano}}, \bibinfo {author}
  {\bibfnamefont {I.}~\bibnamefont {Sagnes}}, \bibinfo {author} {\bibfnamefont
  {A.}~\bibnamefont {Lemaitre}}, \bibinfo {author} {\bibfnamefont {M.~P.}\
  \bibnamefont {Almeida}}, \bibinfo {author} {\bibfnamefont {P.}~\bibnamefont
  {Senellart}},\ and\ \bibinfo {author} {\bibfnamefont {A.~G.}\ \bibnamefont
  {White}},\ }\bibfield  {title} {\bibinfo {title} {Boson sampling with
  single-photon fock states from a bright solid-state source},\ }\href
  {https://doi.org/10.1103/PhysRevLett.118.130503} {\bibfield  {journal}
  {\bibinfo  {journal} {Phys. Rev. Lett.}\ }\textbf {\bibinfo {volume} {118}},\
  \bibinfo {pages} {130503} (\bibinfo {year} {2017})}\BibitemShut {NoStop}%
\bibitem [{\citenamefont {Wang}\ \emph {et~al.}(2017)\citenamefont {Wang},
  \citenamefont {He}, \citenamefont {Li}, \citenamefont {Su}, \citenamefont
  {Li}, \citenamefont {Huang}, \citenamefont {Ding}, \citenamefont {Chen},
  \citenamefont {Liu}, \citenamefont {Qin}, \citenamefont {Li}, \citenamefont
  {He}, \citenamefont {Schneider}, \citenamefont {Kamp}, \citenamefont {Peng},
  \citenamefont {H{\"o}fling}, \citenamefont {Lu},\ and\ \citenamefont
  {Pan}}]{Wang_BS_2017}%
  \BibitemOpen
  \bibfield  {author} {\bibinfo {author} {\bibfnamefont {H.}~\bibnamefont
  {Wang}}, \bibinfo {author} {\bibfnamefont {Y.}~\bibnamefont {He}}, \bibinfo
  {author} {\bibfnamefont {Y.-H.}\ \bibnamefont {Li}}, \bibinfo {author}
  {\bibfnamefont {Z.-E.}\ \bibnamefont {Su}}, \bibinfo {author} {\bibfnamefont
  {B.}~\bibnamefont {Li}}, \bibinfo {author} {\bibfnamefont {H.-L.}\
  \bibnamefont {Huang}}, \bibinfo {author} {\bibfnamefont {X.}~\bibnamefont
  {Ding}}, \bibinfo {author} {\bibfnamefont {M.-C.}\ \bibnamefont {Chen}},
  \bibinfo {author} {\bibfnamefont {C.}~\bibnamefont {Liu}}, \bibinfo {author}
  {\bibfnamefont {J.}~\bibnamefont {Qin}}, \bibinfo {author} {\bibfnamefont
  {J.-P.}\ \bibnamefont {Li}}, \bibinfo {author} {\bibfnamefont {Y.-M.}\
  \bibnamefont {He}}, \bibinfo {author} {\bibfnamefont {C.}~\bibnamefont
  {Schneider}}, \bibinfo {author} {\bibfnamefont {M.}~\bibnamefont {Kamp}},
  \bibinfo {author} {\bibfnamefont {C.-Z.}\ \bibnamefont {Peng}}, \bibinfo
  {author} {\bibfnamefont {S.}~\bibnamefont {H{\"o}fling}}, \bibinfo {author}
  {\bibfnamefont {C.-Y.}\ \bibnamefont {Lu}},\ and\ \bibinfo {author}
  {\bibfnamefont {J.-W.}\ \bibnamefont {Pan}},\ }\bibfield  {title} {\bibinfo
  {title} {High-efficiency multiphoton boson sampling},\ }\href
  {https://doi.org/10.1038/nphoton.2017.63} {\bibfield  {journal} {\bibinfo
  {journal} {Nature Photonics}\ }\textbf {\bibinfo {volume} {11}},\ \bibinfo
  {pages} {361} (\bibinfo {year} {2017})}\BibitemShut {NoStop}%
\bibitem [{\citenamefont {Istrati}\ \emph {et~al.}(2019)\citenamefont
  {Istrati}, \citenamefont {Pilnyak}, \citenamefont {Cohen}, \citenamefont
  {Eisenberg}, \citenamefont {Anton-Solanas}, \citenamefont {Rosillo},
  \citenamefont {Hilaire}, \citenamefont {Ollivier}, \citenamefont {Millet},
  \citenamefont {Lema\'{i}tre}, \citenamefont {Sagnes}, \citenamefont
  {Harouri}, \citenamefont {Lanco},\ and\ \citenamefont
  {Senellart}}]{Istrati_QD_19}%
  \BibitemOpen
  \bibfield  {author} {\bibinfo {author} {\bibfnamefont {D.}~\bibnamefont
  {Istrati}}, \bibinfo {author} {\bibfnamefont {Y.}~\bibnamefont {Pilnyak}},
  \bibinfo {author} {\bibfnamefont {L.}~\bibnamefont {Cohen}}, \bibinfo
  {author} {\bibfnamefont {H.~S.}\ \bibnamefont {Eisenberg}}, \bibinfo {author}
  {\bibfnamefont {C.}~\bibnamefont {Anton-Solanas}}, \bibinfo {author}
  {\bibfnamefont {J.~C.~L.}\ \bibnamefont {Rosillo}}, \bibinfo {author}
  {\bibfnamefont {P.}~\bibnamefont {Hilaire}}, \bibinfo {author} {\bibfnamefont
  {H.}~\bibnamefont {Ollivier}}, \bibinfo {author} {\bibfnamefont
  {C.}~\bibnamefont {Millet}}, \bibinfo {author} {\bibfnamefont
  {A.}~\bibnamefont {Lema\'{i}tre}}, \bibinfo {author} {\bibfnamefont
  {I.}~\bibnamefont {Sagnes}}, \bibinfo {author} {\bibfnamefont
  {A.}~\bibnamefont {Harouri}}, \bibinfo {author} {\bibfnamefont
  {L.}~\bibnamefont {Lanco}},\ and\ \bibinfo {author} {\bibfnamefont
  {P.}~\bibnamefont {Senellart}},\ }\bibfield  {title} {\bibinfo {title}
  {Generating multi-photon entangled states from a single deterministic
  single-photon source},\ }in\ \href {https://doi.org/10.1364/QIM.2019.T3B.1}
  {\emph {\bibinfo {booktitle} {Quantum Information and Measurement (QIM) V:
  Quantum Technologies}}}\ (\bibinfo  {publisher} {Optica Publishing Group},\
  \bibinfo {year} {2019})\ p.\ \bibinfo {pages} {T3B.1}\BibitemShut {NoStop}%
\bibitem [{\citenamefont {Schwartz}\ \emph {et~al.}(2016)\citenamefont
  {Schwartz}, \citenamefont {Cogan}, \citenamefont {Schmidgall}, \citenamefont
  {Don}, \citenamefont {Gantz}, \citenamefont {Kenneth}, \citenamefont
  {Lindner},\ and\ \citenamefont {Gershoni}}]{Schwartz_QD_2016}%
  \BibitemOpen
  \bibfield  {author} {\bibinfo {author} {\bibfnamefont {I.}~\bibnamefont
  {Schwartz}}, \bibinfo {author} {\bibfnamefont {D.}~\bibnamefont {Cogan}},
  \bibinfo {author} {\bibfnamefont {E.~R.}\ \bibnamefont {Schmidgall}},
  \bibinfo {author} {\bibfnamefont {Y.}~\bibnamefont {Don}}, \bibinfo {author}
  {\bibfnamefont {L.}~\bibnamefont {Gantz}}, \bibinfo {author} {\bibfnamefont
  {O.}~\bibnamefont {Kenneth}}, \bibinfo {author} {\bibfnamefont {N.~H.}\
  \bibnamefont {Lindner}},\ and\ \bibinfo {author} {\bibfnamefont
  {D.}~\bibnamefont {Gershoni}},\ }\bibfield  {title} {\bibinfo {title}
  {Deterministic generation of a cluster state of entangled photons},\ }\href
  {https://doi.org/10.1126/science.aah4758} {\bibfield  {journal} {\bibinfo
  {journal} {Science}\ }\textbf {\bibinfo {volume} {354}},\ \bibinfo {pages}
  {434} (\bibinfo {year} {2016})},\ \Eprint
  {https://arxiv.org/abs/https://www.science.org/doi/pdf/10.1126/science.aah4758}
  {https://www.science.org/doi/pdf/10.1126/science.aah4758} \BibitemShut
  {NoStop}%
\bibitem [{\citenamefont {Lu}\ and\ \citenamefont {Pan}(2021)}]{Lu_QD_2021}%
  \BibitemOpen
  \bibfield  {author} {\bibinfo {author} {\bibfnamefont {C.-Y.}\ \bibnamefont
  {Lu}}\ and\ \bibinfo {author} {\bibfnamefont {J.-W.}\ \bibnamefont {Pan}},\
  }\bibfield  {title} {\bibinfo {title} {Quantum-dot single-photon sources for
  the quantum internet},\ }\href {https://doi.org/10.1038/s41565-021-01033-9}
  {\bibfield  {journal} {\bibinfo  {journal} {Nature Nanotechnology}\ }\textbf
  {\bibinfo {volume} {16}},\ \bibinfo {pages} {1294} (\bibinfo {year}
  {2021})}\BibitemShut {NoStop}%
\bibitem [{\citenamefont {Harris}\ \emph {et~al.}(2017)\citenamefont {Harris},
  \citenamefont {Steinbrecher}, \citenamefont {Prabhu}, \citenamefont {Lahini},
  \citenamefont {Mower}, \citenamefont {Bunandar}, \citenamefont {Chen},
  \citenamefont {Wong}, \citenamefont {Baehr-Jones}, \citenamefont {Hochberg},
  \citenamefont {Lloyd},\ and\ \citenamefont {Englund}}]{harris_quantum_2017}%
  \BibitemOpen
  \bibfield  {author} {\bibinfo {author} {\bibfnamefont {N.~C.}\ \bibnamefont
  {Harris}}, \bibinfo {author} {\bibfnamefont {G.~R.}\ \bibnamefont
  {Steinbrecher}}, \bibinfo {author} {\bibfnamefont {M.}~\bibnamefont
  {Prabhu}}, \bibinfo {author} {\bibfnamefont {Y.}~\bibnamefont {Lahini}},
  \bibinfo {author} {\bibfnamefont {J.}~\bibnamefont {Mower}}, \bibinfo
  {author} {\bibfnamefont {D.}~\bibnamefont {Bunandar}}, \bibinfo {author}
  {\bibfnamefont {C.}~\bibnamefont {Chen}}, \bibinfo {author} {\bibfnamefont
  {F.~N.~C.}\ \bibnamefont {Wong}}, \bibinfo {author} {\bibfnamefont
  {T.}~\bibnamefont {Baehr-Jones}}, \bibinfo {author} {\bibfnamefont
  {M.}~\bibnamefont {Hochberg}}, \bibinfo {author} {\bibfnamefont
  {S.}~\bibnamefont {Lloyd}},\ and\ \bibinfo {author} {\bibfnamefont
  {D.}~\bibnamefont {Englund}},\ }\bibfield  {title} {\bibinfo {title} {Quantum
  transport simulations in a programmable nanophotonic processor},\ }\href
  {https://doi.org/10.1038/nphoton.2017.95} {\bibfield  {journal} {\bibinfo
  {journal} {Nature Photonics}\ }\textbf {\bibinfo {volume} {11}},\ \bibinfo
  {pages} {447} (\bibinfo {year} {2017})},\ \bibinfo {note} {number:
  7}\BibitemShut {NoStop}%
\bibitem [{\citenamefont {Taballione}\ \emph {et~al.}(2021)\citenamefont
  {Taballione}, \citenamefont {Meer}, \citenamefont {Snijders}, \citenamefont
  {Hooijschuur}, \citenamefont {Epping}, \citenamefont {Goede}, \citenamefont
  {Kassenberg}, \citenamefont {Venderbosch}, \citenamefont {Toebes},
  \citenamefont {Vlekkert}, \citenamefont {Pinkse},\ and\ \citenamefont
  {Renema}}]{taballione_universal_2021}%
  \BibitemOpen
  \bibfield  {author} {\bibinfo {author} {\bibfnamefont {C.}~\bibnamefont
  {Taballione}}, \bibinfo {author} {\bibfnamefont {R.~v.~d.}\ \bibnamefont
  {Meer}}, \bibinfo {author} {\bibfnamefont {H.~J.}\ \bibnamefont {Snijders}},
  \bibinfo {author} {\bibfnamefont {P.}~\bibnamefont {Hooijschuur}}, \bibinfo
  {author} {\bibfnamefont {J.~P.}\ \bibnamefont {Epping}}, \bibinfo {author}
  {\bibfnamefont {M.~d.}\ \bibnamefont {Goede}}, \bibinfo {author}
  {\bibfnamefont {B.}~\bibnamefont {Kassenberg}}, \bibinfo {author}
  {\bibfnamefont {P.}~\bibnamefont {Venderbosch}}, \bibinfo {author}
  {\bibfnamefont {C.}~\bibnamefont {Toebes}}, \bibinfo {author} {\bibfnamefont
  {H.~v.~d.}\ \bibnamefont {Vlekkert}}, \bibinfo {author} {\bibfnamefont
  {P.~W.~H.}\ \bibnamefont {Pinkse}},\ and\ \bibinfo {author} {\bibfnamefont
  {J.~J.}\ \bibnamefont {Renema}},\ }\bibfield  {title} {\bibinfo {title} {A
  universal fully reconfigurable 12-mode quantum photonic processor},\ }\href
  {https://doi.org/10.1088/2633-4356/ac168c} {\bibfield  {journal} {\bibinfo
  {journal} {Materials for Quantum Technology}\ }\textbf {\bibinfo {volume}
  {1}},\ \bibinfo {pages} {035002} (\bibinfo {year} {2021})},\ \bibinfo {note}
  {publisher: IOP Publishing}\BibitemShut {NoStop}%
\bibitem [{\citenamefont {Taballione}\ \emph {et~al.}(2022)\citenamefont
  {Taballione}, \citenamefont {Anguita}, \citenamefont {de~Goede},
  \citenamefont {Venderbosch}, \citenamefont {Kassenberg}, \citenamefont
  {Snijders}, \citenamefont {Kannan}, \citenamefont {Smith}, \citenamefont
  {Epping}, \citenamefont {van~der Meer}, \citenamefont {Pinkse}, \citenamefont
  {Vlekkert},\ and\ \citenamefont {Renema}}]{Taballione_20mode_2022}%
  \BibitemOpen
  \bibfield  {author} {\bibinfo {author} {\bibfnamefont {C.}~\bibnamefont
  {Taballione}}, \bibinfo {author} {\bibfnamefont {M.~C.}\ \bibnamefont
  {Anguita}}, \bibinfo {author} {\bibfnamefont {M.}~\bibnamefont {de~Goede}},
  \bibinfo {author} {\bibfnamefont {P.}~\bibnamefont {Venderbosch}}, \bibinfo
  {author} {\bibfnamefont {B.}~\bibnamefont {Kassenberg}}, \bibinfo {author}
  {\bibfnamefont {H.}~\bibnamefont {Snijders}}, \bibinfo {author}
  {\bibfnamefont {N.}~\bibnamefont {Kannan}}, \bibinfo {author} {\bibfnamefont
  {D.}~\bibnamefont {Smith}}, \bibinfo {author} {\bibfnamefont {J.~P.}\
  \bibnamefont {Epping}}, \bibinfo {author} {\bibfnamefont {R.}~\bibnamefont
  {van~der Meer}}, \bibinfo {author} {\bibfnamefont {P.~W.~H.}\ \bibnamefont
  {Pinkse}}, \bibinfo {author} {\bibfnamefont {H.~v.~d.}\ \bibnamefont
  {Vlekkert}},\ and\ \bibinfo {author} {\bibfnamefont {J.~J.}\ \bibnamefont
  {Renema}},\ }\href {https://doi.org/10.48550/ARXIV.2203.01801} {\bibinfo
  {title} {20-mode universal quantum photonic processor}} (\bibinfo {year}
  {2022})\BibitemShut {NoStop}%
\bibitem [{\citenamefont {Crespi}\ \emph {et~al.}(2013)\citenamefont {Crespi},
  \citenamefont {Osellame}, \citenamefont {Ramponi}, \citenamefont {Brod},
  \citenamefont {Galvão}, \citenamefont {Spagnolo}, \citenamefont {Vitelli},
  \citenamefont {Maiorino}, \citenamefont {Mataloni},\ and\ \citenamefont
  {Sciarrino}}]{crespi_integrated_2013}%
  \BibitemOpen
  \bibfield  {author} {\bibinfo {author} {\bibfnamefont {A.}~\bibnamefont
  {Crespi}}, \bibinfo {author} {\bibfnamefont {R.}~\bibnamefont {Osellame}},
  \bibinfo {author} {\bibfnamefont {R.}~\bibnamefont {Ramponi}}, \bibinfo
  {author} {\bibfnamefont {D.~J.}\ \bibnamefont {Brod}}, \bibinfo {author}
  {\bibfnamefont {E.~F.}\ \bibnamefont {Galvão}}, \bibinfo {author}
  {\bibfnamefont {N.}~\bibnamefont {Spagnolo}}, \bibinfo {author}
  {\bibfnamefont {C.}~\bibnamefont {Vitelli}}, \bibinfo {author} {\bibfnamefont
  {E.}~\bibnamefont {Maiorino}}, \bibinfo {author} {\bibfnamefont
  {P.}~\bibnamefont {Mataloni}},\ and\ \bibinfo {author} {\bibfnamefont
  {F.}~\bibnamefont {Sciarrino}},\ }\bibfield  {title} {\bibinfo {title}
  {Integrated multimode interferometers with arbitrary designs for photonic
  boson sampling},\ }\href {https://doi.org/10.1038/nphoton.2013.112}
  {\bibfield  {journal} {\bibinfo  {journal} {Nature Photonics}\ }\textbf
  {\bibinfo {volume} {7}},\ \bibinfo {pages} {545} (\bibinfo {year}
  {2013})}\BibitemShut {NoStop}%
\bibitem [{\citenamefont {Dong}\ \emph {et~al.}(2022)\citenamefont {Dong},
  \citenamefont {Clark}, \citenamefont {Leenheer}, \citenamefont {Zimmermann},
  \citenamefont {Dominguez}, \citenamefont {Menssen}, \citenamefont {Heim},
  \citenamefont {Gilbert}, \citenamefont {Englund},\ and\ \citenamefont
  {Eichenfield}}]{Dong_processor_2022}%
  \BibitemOpen
  \bibfield  {author} {\bibinfo {author} {\bibfnamefont {M.}~\bibnamefont
  {Dong}}, \bibinfo {author} {\bibfnamefont {G.}~\bibnamefont {Clark}},
  \bibinfo {author} {\bibfnamefont {A.~J.}\ \bibnamefont {Leenheer}}, \bibinfo
  {author} {\bibfnamefont {M.}~\bibnamefont {Zimmermann}}, \bibinfo {author}
  {\bibfnamefont {D.}~\bibnamefont {Dominguez}}, \bibinfo {author}
  {\bibfnamefont {A.~J.}\ \bibnamefont {Menssen}}, \bibinfo {author}
  {\bibfnamefont {D.}~\bibnamefont {Heim}}, \bibinfo {author} {\bibfnamefont
  {G.}~\bibnamefont {Gilbert}}, \bibinfo {author} {\bibfnamefont
  {D.}~\bibnamefont {Englund}},\ and\ \bibinfo {author} {\bibfnamefont
  {M.}~\bibnamefont {Eichenfield}},\ }\bibfield  {title} {\bibinfo {title}
  {High-speed programmable photonic circuits in a cryogenically compatible,
  visible--near-infrared 200{\thinspace}mm cmos architecture},\ }\href
  {https://doi.org/10.1038/s41566-021-00903-x} {\bibfield  {journal} {\bibinfo
  {journal} {Nature Photonics}\ }\textbf {\bibinfo {volume} {16}},\ \bibinfo
  {pages} {59} (\bibinfo {year} {2022})}\BibitemShut {NoStop}%
\bibitem [{\citenamefont {Hoch}\ \emph {et~al.}(2021)\citenamefont {Hoch},
  \citenamefont {Piacentini}, \citenamefont {Giordani}, \citenamefont {Tian},
  \citenamefont {Iuliano}, \citenamefont {Esposito}, \citenamefont {Camillini},
  \citenamefont {Carvacho}, \citenamefont {Ceccarelli}, \citenamefont
  {Spagnolo}, \citenamefont {Crespi}, \citenamefont {Sciarrino},\ and\
  \citenamefont {Osellame}}]{Hoch_processor_2021}%
  \BibitemOpen
  \bibfield  {author} {\bibinfo {author} {\bibfnamefont {F.}~\bibnamefont
  {Hoch}}, \bibinfo {author} {\bibfnamefont {S.}~\bibnamefont {Piacentini}},
  \bibinfo {author} {\bibfnamefont {T.}~\bibnamefont {Giordani}}, \bibinfo
  {author} {\bibfnamefont {Z.-N.}\ \bibnamefont {Tian}}, \bibinfo {author}
  {\bibfnamefont {M.}~\bibnamefont {Iuliano}}, \bibinfo {author} {\bibfnamefont
  {C.}~\bibnamefont {Esposito}}, \bibinfo {author} {\bibfnamefont
  {A.}~\bibnamefont {Camillini}}, \bibinfo {author} {\bibfnamefont
  {G.}~\bibnamefont {Carvacho}}, \bibinfo {author} {\bibfnamefont
  {F.}~\bibnamefont {Ceccarelli}}, \bibinfo {author} {\bibfnamefont
  {N.}~\bibnamefont {Spagnolo}}, \bibinfo {author} {\bibfnamefont
  {A.}~\bibnamefont {Crespi}}, \bibinfo {author} {\bibfnamefont
  {F.}~\bibnamefont {Sciarrino}},\ and\ \bibinfo {author} {\bibfnamefont
  {R.}~\bibnamefont {Osellame}},\ }\href
  {https://doi.org/10.48550/ARXIV.2106.08260} {\bibinfo {title} {Boson sampling
  in a reconfigurable continuously-coupled 3d photonic circuit}} (\bibinfo
  {year} {2021})\BibitemShut {NoStop}%
\bibitem [{\citenamefont {Tillmann}\ \emph {et~al.}(2013)\citenamefont
  {Tillmann}, \citenamefont {Dakić}, \citenamefont {Heilmann}, \citenamefont
  {Nolte}, \citenamefont {Szameit},\ and\ \citenamefont
  {Walther}}]{tillmann_experimental_2013}%
  \BibitemOpen
  \bibfield  {author} {\bibinfo {author} {\bibfnamefont {M.}~\bibnamefont
  {Tillmann}}, \bibinfo {author} {\bibfnamefont {B.}~\bibnamefont {Dakić}},
  \bibinfo {author} {\bibfnamefont {R.}~\bibnamefont {Heilmann}}, \bibinfo
  {author} {\bibfnamefont {S.}~\bibnamefont {Nolte}}, \bibinfo {author}
  {\bibfnamefont {A.}~\bibnamefont {Szameit}},\ and\ \bibinfo {author}
  {\bibfnamefont {P.}~\bibnamefont {Walther}},\ }\bibfield  {title} {\bibinfo
  {title} {Experimental boson sampling},\ }\href
  {https://doi.org/10.1038/nphoton.2013.102} {\bibfield  {journal} {\bibinfo
  {journal} {Nature Photonics}\ }\textbf {\bibinfo {volume} {7}},\ \bibinfo
  {pages} {540} (\bibinfo {year} {2013})},\ \bibinfo {note} {number:
  7}\BibitemShut {NoStop}%
\bibitem [{\citenamefont {Spring}\ \emph {et~al.}(2013)\citenamefont {Spring},
  \citenamefont {Metcalf}, \citenamefont {Humphreys}, \citenamefont
  {Kolthammer}, \citenamefont {Jin}, \citenamefont {Barbieri}, \citenamefont
  {Datta}, \citenamefont {Thomas-Peter}, \citenamefont {Langford},
  \citenamefont {Kundys}, \citenamefont {Gates}, \citenamefont {Smith},
  \citenamefont {Smith},\ and\ \citenamefont {Walmsley}}]{spring_boson_2013}%
  \BibitemOpen
  \bibfield  {author} {\bibinfo {author} {\bibfnamefont {J.~B.}\ \bibnamefont
  {Spring}}, \bibinfo {author} {\bibfnamefont {B.~J.}\ \bibnamefont {Metcalf}},
  \bibinfo {author} {\bibfnamefont {P.~C.}\ \bibnamefont {Humphreys}}, \bibinfo
  {author} {\bibfnamefont {W.~S.}\ \bibnamefont {Kolthammer}}, \bibinfo
  {author} {\bibfnamefont {X.-M.}\ \bibnamefont {Jin}}, \bibinfo {author}
  {\bibfnamefont {M.}~\bibnamefont {Barbieri}}, \bibinfo {author}
  {\bibfnamefont {A.}~\bibnamefont {Datta}}, \bibinfo {author} {\bibfnamefont
  {N.}~\bibnamefont {Thomas-Peter}}, \bibinfo {author} {\bibfnamefont {N.~K.}\
  \bibnamefont {Langford}}, \bibinfo {author} {\bibfnamefont {D.}~\bibnamefont
  {Kundys}}, \bibinfo {author} {\bibfnamefont {J.~C.}\ \bibnamefont {Gates}},
  \bibinfo {author} {\bibfnamefont {B.~J.}\ \bibnamefont {Smith}}, \bibinfo
  {author} {\bibfnamefont {P.~G.~R.}\ \bibnamefont {Smith}},\ and\ \bibinfo
  {author} {\bibfnamefont {I.~A.}\ \bibnamefont {Walmsley}},\ }\bibfield
  {title} {\bibinfo {title} {Boson {Sampling} on a {Photonic} {Chip}},\ }\href
  {https://doi.org/10.1126/science.1231692} {\bibfield  {journal} {\bibinfo
  {journal} {Science}\ }\textbf {\bibinfo {volume} {339}},\ \bibinfo {pages}
  {798} (\bibinfo {year} {2013})},\ \bibinfo {note} {number: 6121}\BibitemShut
  {NoStop}%
\bibitem [{\citenamefont {Carolan}\ \emph {et~al.}(2015)\citenamefont
  {Carolan}, \citenamefont {Harrold}, \citenamefont {Sparrow}, \citenamefont
  {Martín-López}, \citenamefont {Russell}, \citenamefont {Silverstone},
  \citenamefont {Shadbolt}, \citenamefont {Matsuda}, \citenamefont {Oguma},
  \citenamefont {Itoh}, \citenamefont {Marshall}, \citenamefont {Thompson},
  \citenamefont {Matthews}, \citenamefont {Hashimoto}, \citenamefont
  {O’Brien},\ and\ \citenamefont {Laing}}]{carolan_universal_2015}%
  \BibitemOpen
  \bibfield  {author} {\bibinfo {author} {\bibfnamefont {J.}~\bibnamefont
  {Carolan}}, \bibinfo {author} {\bibfnamefont {C.}~\bibnamefont {Harrold}},
  \bibinfo {author} {\bibfnamefont {C.}~\bibnamefont {Sparrow}}, \bibinfo
  {author} {\bibfnamefont {E.}~\bibnamefont {Martín-López}}, \bibinfo
  {author} {\bibfnamefont {N.~J.}\ \bibnamefont {Russell}}, \bibinfo {author}
  {\bibfnamefont {J.~W.}\ \bibnamefont {Silverstone}}, \bibinfo {author}
  {\bibfnamefont {P.~J.}\ \bibnamefont {Shadbolt}}, \bibinfo {author}
  {\bibfnamefont {N.}~\bibnamefont {Matsuda}}, \bibinfo {author} {\bibfnamefont
  {M.}~\bibnamefont {Oguma}}, \bibinfo {author} {\bibfnamefont
  {M.}~\bibnamefont {Itoh}}, \bibinfo {author} {\bibfnamefont {G.~D.}\
  \bibnamefont {Marshall}}, \bibinfo {author} {\bibfnamefont {M.~G.}\
  \bibnamefont {Thompson}}, \bibinfo {author} {\bibfnamefont {J.~C.~F.}\
  \bibnamefont {Matthews}}, \bibinfo {author} {\bibfnamefont {T.}~\bibnamefont
  {Hashimoto}}, \bibinfo {author} {\bibfnamefont {J.~L.}\ \bibnamefont
  {O’Brien}},\ and\ \bibinfo {author} {\bibfnamefont {A.}~\bibnamefont
  {Laing}},\ }\bibfield  {title} {\bibinfo {title} {Universal linear optics},\
  }\href {https://doi.org/10.1126/science.aab3642} {\bibfield  {journal}
  {\bibinfo  {journal} {Science}\ }\textbf {\bibinfo {volume} {349}},\ \bibinfo
  {pages} {711} (\bibinfo {year} {2015})},\ \bibinfo {note} {number:
  6249}\BibitemShut {NoStop}%
\bibitem [{\citenamefont {Mennea}\ \emph {et~al.}(2018)\citenamefont {Mennea},
  \citenamefont {Clements}, \citenamefont {Smith}, \citenamefont {Gates},
  \citenamefont {Metcalf}, \citenamefont {Bannerman}, \citenamefont {Burgwal},
  \citenamefont {Renema}, \citenamefont {Kolthammer}, \citenamefont
  {Walmsley},\ and\ \citenamefont {Smith}}]{mennea_modular_2018}%
  \BibitemOpen
  \bibfield  {author} {\bibinfo {author} {\bibfnamefont {P.~L.}\ \bibnamefont
  {Mennea}}, \bibinfo {author} {\bibfnamefont {W.~R.}\ \bibnamefont
  {Clements}}, \bibinfo {author} {\bibfnamefont {D.~H.}\ \bibnamefont {Smith}},
  \bibinfo {author} {\bibfnamefont {J.~C.}\ \bibnamefont {Gates}}, \bibinfo
  {author} {\bibfnamefont {B.~J.}\ \bibnamefont {Metcalf}}, \bibinfo {author}
  {\bibfnamefont {R.~H.~S.}\ \bibnamefont {Bannerman}}, \bibinfo {author}
  {\bibfnamefont {R.}~\bibnamefont {Burgwal}}, \bibinfo {author} {\bibfnamefont
  {J.~J.}\ \bibnamefont {Renema}}, \bibinfo {author} {\bibfnamefont {W.~S.}\
  \bibnamefont {Kolthammer}}, \bibinfo {author} {\bibfnamefont {I.~A.}\
  \bibnamefont {Walmsley}},\ and\ \bibinfo {author} {\bibfnamefont {P.~G.~R.}\
  \bibnamefont {Smith}},\ }\bibfield  {title} {\bibinfo {title} {Modular linear
  optical circuits},\ }\href {https://doi.org/10.1364/OPTICA.5.001087}
  {\bibfield  {journal} {\bibinfo  {journal} {Optica}\ }\textbf {\bibinfo
  {volume} {5}},\ \bibinfo {pages} {1087} (\bibinfo {year} {2018})},\ \bibinfo
  {note} {publisher: OSA}\BibitemShut {NoStop}%
\bibitem [{\citenamefont {Taballione}\ \emph {et~al.}(2019)\citenamefont
  {Taballione}, \citenamefont {Wolterink}, \citenamefont {Lugani},
  \citenamefont {Eckstein}, \citenamefont {Bell}, \citenamefont {Grootjans},
  \citenamefont {Visscher}, \citenamefont {Geskus}, \citenamefont {Roeloffzen},
  \citenamefont {Renema}, \citenamefont {Walmsley}, \citenamefont {Pinkse},\
  and\ \citenamefont {Boller}}]{taballione_88_2019}%
  \BibitemOpen
  \bibfield  {author} {\bibinfo {author} {\bibfnamefont {C.}~\bibnamefont
  {Taballione}}, \bibinfo {author} {\bibfnamefont {T.~A.~W.}\ \bibnamefont
  {Wolterink}}, \bibinfo {author} {\bibfnamefont {J.}~\bibnamefont {Lugani}},
  \bibinfo {author} {\bibfnamefont {A.}~\bibnamefont {Eckstein}}, \bibinfo
  {author} {\bibfnamefont {B.~A.}\ \bibnamefont {Bell}}, \bibinfo {author}
  {\bibfnamefont {R.}~\bibnamefont {Grootjans}}, \bibinfo {author}
  {\bibfnamefont {I.}~\bibnamefont {Visscher}}, \bibinfo {author}
  {\bibfnamefont {D.}~\bibnamefont {Geskus}}, \bibinfo {author} {\bibfnamefont
  {C.~G.~H.}\ \bibnamefont {Roeloffzen}}, \bibinfo {author} {\bibfnamefont
  {J.~J.}\ \bibnamefont {Renema}}, \bibinfo {author} {\bibfnamefont {I.~A.}\
  \bibnamefont {Walmsley}}, \bibinfo {author} {\bibfnamefont {P.~W.~H.}\
  \bibnamefont {Pinkse}},\ and\ \bibinfo {author} {\bibfnamefont {K.-J.}\
  \bibnamefont {Boller}},\ }\bibfield  {title} {\bibinfo {title} {8×8
  reconfigurable quantum photonic processor based on silicon nitride
  waveguides},\ }\href {https://doi.org/10.1364/OE.27.026842} {\bibfield
  {journal} {\bibinfo  {journal} {Optics Express}\ }\textbf {\bibinfo {volume}
  {27}},\ \bibinfo {pages} {26842} (\bibinfo {year} {2019})},\ \bibinfo {note}
  {number: 19 Publisher: OSA}\BibitemShut {NoStop}%
\bibitem [{\citenamefont {{C. G. H. Roeloffzen}}\ \emph
  {et~al.}(2018)\citenamefont {{C. G. H. Roeloffzen}}, \citenamefont {{M.
  Hoekman}}, \citenamefont {{E. J. Klein}}, \citenamefont {{L. S. Wevers}},
  \citenamefont {{R. B. Timens}}, \citenamefont {{D. Marchenko}}, \citenamefont
  {{D. Geskus}}, \citenamefont {{R. Dekker}}, \citenamefont {{A. Alippi}},
  \citenamefont {{R. Grootjans}}, \citenamefont {{A. van Rees}}, \citenamefont
  {{R. M. Oldenbeuving}}, \citenamefont {{J. P. Epping}}, \citenamefont {{R. G.
  Heideman}}, \citenamefont {{K. Wörhoff}}, \citenamefont {{A. Leinse}},
  \citenamefont {{D. Geuzebroek}}, \citenamefont {{E. Schreuder}},
  \citenamefont {{P. W. L. van Dijk}}, \citenamefont {{I. Visscher}},
  \citenamefont {{C. Taddei}}, \citenamefont {{Y. Fan}}, \citenamefont {{C.
  Taballione}}, \citenamefont {{Y. Liu}}, \citenamefont {{D. Marpaung}},
  \citenamefont {{L. Zhuang}}, \citenamefont {{M. Benelajla}},\ and\
  \citenamefont {{K. Boller}}}]{c_g_h_roeloffzen_low-loss_2018}%
  \BibitemOpen
  \bibfield  {author} {\bibinfo {author} {\bibnamefont {{C. G. H.
  Roeloffzen}}}, \bibinfo {author} {\bibnamefont {{M. Hoekman}}}, \bibinfo
  {author} {\bibnamefont {{E. J. Klein}}}, \bibinfo {author} {\bibnamefont {{L.
  S. Wevers}}}, \bibinfo {author} {\bibnamefont {{R. B. Timens}}}, \bibinfo
  {author} {\bibnamefont {{D. Marchenko}}}, \bibinfo {author} {\bibnamefont
  {{D. Geskus}}}, \bibinfo {author} {\bibnamefont {{R. Dekker}}}, \bibinfo
  {author} {\bibnamefont {{A. Alippi}}}, \bibinfo {author} {\bibnamefont {{R.
  Grootjans}}}, \bibinfo {author} {\bibnamefont {{A. van Rees}}}, \bibinfo
  {author} {\bibnamefont {{R. M. Oldenbeuving}}}, \bibinfo {author}
  {\bibnamefont {{J. P. Epping}}}, \bibinfo {author} {\bibnamefont {{R. G.
  Heideman}}}, \bibinfo {author} {\bibnamefont {{K. Wörhoff}}}, \bibinfo
  {author} {\bibnamefont {{A. Leinse}}}, \bibinfo {author} {\bibnamefont {{D.
  Geuzebroek}}}, \bibinfo {author} {\bibnamefont {{E. Schreuder}}}, \bibinfo
  {author} {\bibnamefont {{P. W. L. van Dijk}}}, \bibinfo {author}
  {\bibnamefont {{I. Visscher}}}, \bibinfo {author} {\bibnamefont {{C.
  Taddei}}}, \bibinfo {author} {\bibnamefont {{Y. Fan}}}, \bibinfo {author}
  {\bibnamefont {{C. Taballione}}}, \bibinfo {author} {\bibnamefont {{Y.
  Liu}}}, \bibinfo {author} {\bibnamefont {{D. Marpaung}}}, \bibinfo {author}
  {\bibnamefont {{L. Zhuang}}}, \bibinfo {author} {\bibnamefont {{M.
  Benelajla}}},\ and\ \bibinfo {author} {\bibnamefont {{K. Boller}}},\
  }\bibfield  {title} {\bibinfo {title} {Low-{Loss} {Si3N4} {TriPleX} {Optical}
  {Waveguides}: {Technology} and {Applications} {Overview}},\ }\href
  {https://doi.org/10.1109/JSTQE.2018.2793945} {\bibfield  {journal} {\bibinfo
  {journal} {IEEE Journal of Selected Topics in Quantum Electronics}\ }\textbf
  {\bibinfo {volume} {24}},\ \bibinfo {pages} {1} (\bibinfo {year} {2018})},\
  \bibinfo {note} {number: 4}\BibitemShut {NoStop}%
\bibitem [{\citenamefont {Eisaman}\ \emph {et~al.}(2011)\citenamefont
  {Eisaman}, \citenamefont {Fan}, \citenamefont {Migdall},\ and\ \citenamefont
  {Polyakov}}]{eisaman2011invited}%
  \BibitemOpen
  \bibfield  {author} {\bibinfo {author} {\bibfnamefont {M.~D.}\ \bibnamefont
  {Eisaman}}, \bibinfo {author} {\bibfnamefont {J.}~\bibnamefont {Fan}},
  \bibinfo {author} {\bibfnamefont {A.}~\bibnamefont {Migdall}},\ and\ \bibinfo
  {author} {\bibfnamefont {S.~V.}\ \bibnamefont {Polyakov}},\ }\bibfield
  {title} {\bibinfo {title} {Invited review article: Single-photon sources and
  detectors},\ }\href@noop {} {\bibfield  {journal} {\bibinfo  {journal}
  {Review of scientific instruments}\ }\textbf {\bibinfo {volume} {82}},\
  \bibinfo {pages} {071101} (\bibinfo {year} {2011})}\BibitemShut {NoStop}%
\bibitem [{\citenamefont {Clements}\ \emph {et~al.}(2016)\citenamefont
  {Clements}, \citenamefont {Humphreys}, \citenamefont {Metcalf}, \citenamefont
  {Kolthammer},\ and\ \citenamefont {Walmsley}}]{clements_optimal_2016}%
  \BibitemOpen
  \bibfield  {author} {\bibinfo {author} {\bibfnamefont {W.~R.}\ \bibnamefont
  {Clements}}, \bibinfo {author} {\bibfnamefont {P.~C.}\ \bibnamefont
  {Humphreys}}, \bibinfo {author} {\bibfnamefont {B.~J.}\ \bibnamefont
  {Metcalf}}, \bibinfo {author} {\bibfnamefont {W.~S.}\ \bibnamefont
  {Kolthammer}},\ and\ \bibinfo {author} {\bibfnamefont {I.~A.}\ \bibnamefont
  {Walmsley}},\ }\bibfield  {title} {\bibinfo {title} {Optimal design for
  universal multiport interferometers},\ }\href
  {https://doi.org/10.1364/OPTICA.3.001460} {\bibfield  {journal} {\bibinfo
  {journal} {Optica}\ }\textbf {\bibinfo {volume} {3}},\ \bibinfo {pages}
  {1460} (\bibinfo {year} {2016})},\ \bibinfo {note} {number: 12 Publisher:
  OSA}\BibitemShut {NoStop}%
\bibitem [{\citenamefont {Epping}\ \emph {et~al.}(2017)\citenamefont {Epping},
  \citenamefont {Marchenko}, \citenamefont {Leinse}, \citenamefont {Mateman},
  \citenamefont {Hoekman}, \citenamefont {Wevers}, \citenamefont {Klein},
  \citenamefont {Roeloffzen}, \citenamefont {Dekkers},\ and\ \citenamefont
  {Heilmann}}]{epping_ultra-low-power_2017}%
  \BibitemOpen
  \bibfield  {author} {\bibinfo {author} {\bibfnamefont {J.~P.}\ \bibnamefont
  {Epping}}, \bibinfo {author} {\bibfnamefont {D.}~\bibnamefont {Marchenko}},
  \bibinfo {author} {\bibfnamefont {A.}~\bibnamefont {Leinse}}, \bibinfo
  {author} {\bibfnamefont {R.}~\bibnamefont {Mateman}}, \bibinfo {author}
  {\bibfnamefont {M.}~\bibnamefont {Hoekman}}, \bibinfo {author} {\bibfnamefont
  {L.}~\bibnamefont {Wevers}}, \bibinfo {author} {\bibfnamefont {E.~J.}\
  \bibnamefont {Klein}}, \bibinfo {author} {\bibfnamefont {C.~G.~H.}\
  \bibnamefont {Roeloffzen}}, \bibinfo {author} {\bibfnamefont
  {M.}~\bibnamefont {Dekkers}},\ and\ \bibinfo {author} {\bibfnamefont
  {R.}~\bibnamefont {Heilmann}},\ }\bibfield  {title} {\bibinfo {title}
  {Ultra-low-power stress-based phase actuator for microwave photonics},\
  }\href
  {http://www.osapublishing.org/abstract.cfm?URI=CLEO_Europe-2017-CK_7_6}
  {\bibfield  {journal} {\bibinfo  {journal} {2017 European Conference on
  Lasers and Electro-Optics and European Quantum Electronics Conference}\
  }\textbf {\bibinfo {volume} {CK\_7\_6}} (\bibinfo {year} {2017})},\ \bibinfo
  {note} {journal Abbreviation: CLEO\_Europe}\BibitemShut {NoStop}%
\bibitem [{\citenamefont {Bauters}\ \emph {et~al.}(2011)\citenamefont
  {Bauters}, \citenamefont {Heck}, \citenamefont {John}, \citenamefont
  {Barton}, \citenamefont {Bruinink}, \citenamefont {Leinse}, \citenamefont
  {Heideman}, \citenamefont {Blumenthal},\ and\ \citenamefont
  {Bowers}}]{bauters_planar_2011}%
  \BibitemOpen
  \bibfield  {author} {\bibinfo {author} {\bibfnamefont {J.~F.}\ \bibnamefont
  {Bauters}}, \bibinfo {author} {\bibfnamefont {M.~J.~R.}\ \bibnamefont
  {Heck}}, \bibinfo {author} {\bibfnamefont {D.~D.}\ \bibnamefont {John}},
  \bibinfo {author} {\bibfnamefont {J.~S.}\ \bibnamefont {Barton}}, \bibinfo
  {author} {\bibfnamefont {C.~M.}\ \bibnamefont {Bruinink}}, \bibinfo {author}
  {\bibfnamefont {A.}~\bibnamefont {Leinse}}, \bibinfo {author} {\bibfnamefont
  {R.~G.}\ \bibnamefont {Heideman}}, \bibinfo {author} {\bibfnamefont {D.~J.}\
  \bibnamefont {Blumenthal}},\ and\ \bibinfo {author} {\bibfnamefont {J.~E.}\
  \bibnamefont {Bowers}},\ }\bibfield  {title} {\bibinfo {title} {Planar
  waveguides with less than 0.1 {dB}/m propagation loss fabricated with wafer
  bonding},\ }\href {https://doi.org/10.1364/OE.19.024090} {\bibfield
  {journal} {\bibinfo  {journal} {Opt. Express}\ }\textbf {\bibinfo {volume}
  {19}},\ \bibinfo {pages} {24090} (\bibinfo {year} {2011})},\ \bibinfo {note}
  {publisher: OSA}\BibitemShut {NoStop}%
\bibitem [{\citenamefont {Schuck}\ \emph {et~al.}(2013)\citenamefont {Schuck},
  \citenamefont {Pernice},\ and\ \citenamefont {Tang}}]{schuck_nbtin_2013}%
  \BibitemOpen
  \bibfield  {author} {\bibinfo {author} {\bibfnamefont {C.}~\bibnamefont
  {Schuck}}, \bibinfo {author} {\bibfnamefont {W.~H.~P.}\ \bibnamefont
  {Pernice}},\ and\ \bibinfo {author} {\bibfnamefont {H.~X.}\ \bibnamefont
  {Tang}},\ }\bibfield  {title} {\bibinfo {title} {{NbTiN} superconducting
  nanowire detectors for visible and telecom wavelengths single photon counting
  on {Si3N4} photonic circuits},\ }\href {https://doi.org/10.1063/1.4788931}
  {\bibfield  {journal} {\bibinfo  {journal} {Applied Physics Letters}\
  }\textbf {\bibinfo {volume} {102}},\ \bibinfo {pages} {051101} (\bibinfo
  {year} {2013})},\ \bibinfo {note} {number: 5 Publisher: American Institute of
  Physics}\BibitemShut {NoStop}%
\bibitem [{\citenamefont {Schuck}\ \emph {et~al.}(2016)\citenamefont {Schuck},
  \citenamefont {Guo}, \citenamefont {Fan}, \citenamefont {Ma}, \citenamefont
  {Poot},\ and\ \citenamefont {Tang}}]{schuck_quantum_2016}%
  \BibitemOpen
  \bibfield  {author} {\bibinfo {author} {\bibfnamefont {C.}~\bibnamefont
  {Schuck}}, \bibinfo {author} {\bibfnamefont {X.}~\bibnamefont {Guo}},
  \bibinfo {author} {\bibfnamefont {L.}~\bibnamefont {Fan}}, \bibinfo {author}
  {\bibfnamefont {X.}~\bibnamefont {Ma}}, \bibinfo {author} {\bibfnamefont
  {M.}~\bibnamefont {Poot}},\ and\ \bibinfo {author} {\bibfnamefont {H.~X.}\
  \bibnamefont {Tang}},\ }\bibfield  {title} {\bibinfo {title} {Quantum
  interference in heterogeneous superconducting-photonic circuits on a silicon
  chip},\ }\href {https://doi.org/10.1038/ncomms10352} {\bibfield  {journal}
  {\bibinfo  {journal} {Nature Communications}\ }\textbf {\bibinfo {volume}
  {7}},\ \bibinfo {pages} {10352} (\bibinfo {year} {2016})},\ \bibinfo {note}
  {number: 1}\BibitemShut {NoStop}%
\end{thebibliography}%

\end{document}